\documentclass[sigconf]{acmart}
\usepackage{graphicx} % Required for inserting images
\usepackage{subcaption}

\copyrightyear{2025}
\acmYear{2025}
\setcopyright{rightsretained}
\acmConference[CHI EA '25]{Extended Abstracts of the CHI Conference on Human Factors in Computing Systems}{April 26-May 1, 2025}{Yokohama, Japan}
\acmBooktitle{Extended Abstracts of the CHI Conference on Human Factors in Computing Systems (CHI EA '25), April 26-May 1, 2025, Yokohama, Japan}\acmDOI{10.1145/3706599.3720063}
\acmISBN{979-8-4007-1395-8/2025/04}

\begin{document}

\title[Unveiling the Scarce Contributions of UX Professionals to Usability Issue Discussions of OSS Projects]{Untold Stories: Unveiling the Scarce Contributions of UX Professionals to Usability Issue Discussions of Open Source Software Projects}

\author{Arghavan Sanei}
\email{arghavan.sanei@polymtl.ca}
\affiliation{%
  \institution{Polytechnique Montreal}
  \city{Montreal}
  \state{Quebec}
  \country{Canada}
}
\author{Jinghui Cheng}
\email{Jinghui.cheng@polymtl.ca}
\affiliation{%
  \institution{Polytechnique Montreal}
  \city{Montreal}
  \state{Quebec}
  \country{Canada}
}

\begin{abstract}
Previous work established that open source software (OSS) projects can benefit from the involvement of UX professionals, who offer user-centric perspectives and contributions to improve software usability. However, their participation in OSS issue discussions (places where design and implementation decisions are often made) is relatively scarce since those platforms are created with a developer-centric mindset. Analyzing a dataset sampled from five OSS projects, this study identifies UX professionals' distinct approaches to raising and following up on usability issues. Compared to other contributors, UX professionals addressed a broader range of usability issues, well-supported their stances, and were more factual than emotional. They also actively engage in discussions to provide additional insights and clarifications in comments following up on the issues they posted. Results from this study provide useful insights for increasing UX professionals' involvement in OSS communities to improve usability and end-user satisfaction.
\end{abstract}

\begin{CCSXML}
<ccs2012>
<concept>
<concept_id>10003120.10003130.10003233.10003597</concept_id>
<concept_desc>Human-centered computing~Open source software</concept_desc>
<concept_significance>300</concept_significance>
</concept>
<concept>
<concept_id>10011007.10010940.10011003.10011687</concept_id>
<concept_desc>Software and its engineering~Software usability</concept_desc>
<concept_significance>500</concept_significance>
</concept>
<concept>
<concept_id>10003120.10003130.10003131.10003570</concept_id>
<concept_desc>Human-centered computing~Computer supported cooperative work</concept_desc>
<concept_significance>300</concept_significance>
</concept>
</ccs2012>
\end{CCSXML}

\ccsdesc[500]{Human-centered computing~Open source software}
\ccsdesc[500]{Software and its engineering~Software usability}
\ccsdesc[500]{Human-centered computing~Computer supported cooperative work}

\keywords{Usability Discussions, Open Source Software, UX Professionals}

\maketitle

\section{Introduction}
Open-source software (OSS) relies on the distributed, often asynchronous, collaborative efforts of developers, designers, and users to improve usability~\cite{wang2020open, Cheng2018, Hellman2022}. This collaboration frequently happens in issue tracking systems (ITSs), such as GitHub Issues~\cite{githubGitHubIssues}, in which various bugs, enhancement requests, and other project-related topics were raised, discussed, and traced. These topics frequently touched on usability aspects of the product~\cite{sanei2023characterizing}. Previous studies on usability discussions on the ITSs~\cite{Cheng2018, sanei2023characterizing} indicated that contributors with UX backgrounds may play a crucial role in enhancing the usability of OSS, providing reports, feedback, and insights that contribute to user-centered improvements. In this study, we investigate the impacts of UX professionals and their expertise in collaborating with the OSS community within the ITS. 

While many contributors may raise and discuss usability issues, individuals with UX design backgrounds bring a unique perspective. They do not only identify problems but also offer solutions rooted in user-centric design principles~\cite{ccetin2007analysis}. Moreover, utilizing their knowledge, UX experts' inputs can go beyond identifying existing bugs and problems in OSS and push into anticipating and preventing potential user difficulties and improving user satisfaction. Recognizing and encouraging the active collaboration and engagement of UX experts within the larger OSS communities can significantly elevate the overall usability and quality of OSS~\cite{hedberg2009integrating, khalajzadeh2022diverse, Rajanen2023Usability}. 

However, previous exploration~\cite{sanei2023characterizing} indicated that the participation of contributors who have UX backgrounds in issue discussions may not be very frequent. This is partially because ITSs are often created with developers in mind, ignoring the needs of other stakeholders, such as end users and designers. Although scarce, UX professionals' participation in issue discussions can still reveal important insights about how they, and their expertise, can influence the style of their contribution in a developer-centric environment, which subsequently impacts the OSS community and the software design. Their participation in ITSs is also crucial for their voices to be heard since design and implementation decisions are often made on those platforms. By analyzing their contribution, we hope to highlight the necessity for their design expertise in the OSS community, emphasizing the potential opportunities for creating more inclusive OSS tools and development environments.

In particular, we conducted a mixed-methods analysis on usability issue discussions in five widely used OSS projects (Jupyter Lab, Google Colab, CoCalc, VSCode, and Atom), based on a dataset created in previous study~\cite{sanei2023characterizing}. We first focused on discovering the background and experience of usability issue posters to identify UX professionals from the dataset. After we found the UX professionals, we explored the following two research questions to characterize their contributions to the projects.

\textbf{RQ1: How do UX professionals raise usability issues differently than other contributors?}
Although there are studies that focus on usability issue discussions in OSS development~\cite{arora2013state, garcia2017challenges, sharma2018usability, sanei2023characterizing}, there is no research in investigating issues posted by UX professionals in OSS. To address this gap, we conducted an analysis to identify the aspects of usability (based on Nielsen's heuristics~\cite{nielsen2005ten}) that have been reported by UX professionals, the types of sentiment and tone expressed in those reports, and the argumentative discourse embedded in the posted usability issues. The findings indicated that there is a limited number of UX professionals in projects who participated in issue discussions (around one in each project). Compared with other participants, UX professionals considered a wider spectrum of usability issues, sufficiently supported their stances when reporting issues, and reported issues mostly based on facts, not their emotions. 

\textbf{RQ2: How do UX professionals follow up on the usability issues they posted?}
After characterizing the issues reported by the UX professionals, we investigated how they continued to contribute to resolving those issues they posted. Concretely, we analyzed their behavior after posting usability issues, when following up on those issues in comments. Through this analysis, we found that UX professionals engaged in about 1/3 of their issues as commenters. An inductive coding on the purpose of their follow-up comments indicated that their main goals were to share ideas/opinions/experiences and provide supplementary details so that their feedback or suggestions could be better understood, taken seriously, and addressed by the OSS developers.

In summary, this study concentrated on examining how UX professionals participated in OSS ITSs, exploring their contribution to reporting and following up on usability issues. Overall, this investigation revealed the distinctive contribution but minimal participation of UX professionals in addressing usability within the ITS. Our results provided some insights that can inform strategies to encourage their active involvement, leading to enriching OSS communities, fostering OSS usability quality, and improving end-user satisfaction.
\section{Methods}
We based our analyses on the labeled data created in previous work~\cite{sanei2023characterizing}. The dataset distinguished 305 usability issues from five popular OSS projects (Jupyter Lab,
Google Colab, CoCalc, VSCode, and Atom) and identified their posters. In this paper, we focus on individuals who have ever posted a usability issue in that dataset. 

\subsection{Discovering the Role of Issue Posters}\label{sec: Discovering_role}

To detect the background of the usability issue posters in the dataset, we checked each user's \textit{Profile page} on GitHub, examining their bios, shared personal websites, LinkedIn pages, and/or shared resumes. If they have not shared these information, we searched for their LinkedIn profiles using their full names to extract details on their backgrounds and expertise. We considered their job titles posted in the information acquired this way and categorized them into (1) UX professionals, (2) managers, (3) data scientists, and (4) developers. UX professionals were defined as those indicating positions such as \textit{UX designer} and \textit{user interface and user experience designer}.

Among the 224 usability issue posters in the dataset, we were able to identify the role of 180 users. Within those 180 users, 121 (67.2\%) were developers, 34 (18.9\%) identified as data scientists, 21 (11.7\%) held managerial positions, and only four (2.2\%) were UX professionals. The UX professionals included one male contributed to \textit{VSCode}, another male contributed to \textit{Atom}, and two involved in \textit{Jupyter Lab} project, one male and one female. Notably, there were no UX professionals involved in \textit{CoCalc} and \textit{Google Colab} projects in our data sample. For easier referencing, in the following we call the UX professionals of VSCode as \textit{VSCode\_pro}, Atom \textit{Atom\_pro}, male of Jupyter Lab as \textit{Jupyter\_pro\_M} and female as \textit{Jupyter\_pro\_F}.

\subsection{Characteristics of Issues Posted by UX Professionals (RQ1)}

Once we identified the roles of the usability issue posters, we extracted all the issues posted by the four UX professionals across the five OSS projects. Next, we analyzed the extracted issues by adopting the following steps. First, following the approach outlined in \cite{sanei2023characterizing}, we labeled each issue with either usability or non-usability; and for each usability issue, we identified the main \textit{usability dimension} touched by the issue using the ten Nielsen heuristics~\cite{nielsen2005ten}. Then, similar to \cite{sanei2021impacts}, we identified the specific \textit{sentiment} and \textit{tone} expressed by the UX professionals when posting the usability issues. In our study, the sentiment captures the valence of the emotion that includes three categories (positive, negative, and neutral), while the tone describes emotion with seven affective factors (excited, frustrated, impolite, polite, sad, satisfied, and sympathetic). Subsequently, we analyzed the \textit{argument structure} of the usability issues to better understand the discursive device that the issue posters adopted to convince other discussion participants. We particularly identified whether a \textit{claim} and a \textit{premise} appeared in a usability issue post, using criteria proposed in prior work~\cite{skitalinskaya_learning_2021, wachsmuth_argumentation_2017, dowden1993logical}. Statements were considered as claims if they explicitly indicate the position or stance of the issue posters to the discussed usability issues; and premise means that a statement contains reasoning, evidence, or examples that support a stance. We compared how the above characteristics (i.e., usability dimensions, sentiments, tones, and argument structures) differed in issues posted by UX professionals and those without UX expertise.

\subsection{UX Professionals' Purpose Following Up on Issues (RQ2)}

% After investigating how UX professionals posted the usability issues, we recognized the importance of understanding their participation afterwards, particularly in following up on the discussion threads of the issues they posted. 
Thus, we first isolated comments made by the UX professionals posted to the usability issues they created within the datasets. Then, we employed an inductive content analysis~\cite{wamboldt1992content, Hsieh2005} and categorized the various purposes behind their contributions in posting each comment. For our analysis, the \textit{purpose} specifies the distinct goal that a particular comment serves within the context of the discussion thread. The purpose of a comment may vary based on its content and the immediate objective of the issue posters to write in the discussion to address one specific comment posted by another contributor. We grouped the identified purposes into themes through an iterative approach conducted by the two authors.

\section{Results}
In our analysis, we found 105 issues reported by the four UX professionals in all the issues of VSCode, Atom, and Jupyter Lab (a total of 139,948 issues). A majority of the issues (93 out of 105, or 88.6\%) focused on usability, only one (0.9\%) discussed a bug and 11 (10.5\%) were related to reporting \textit{UX meetings} for the VSCode project.
\subsection{RQ1: How Do UX Professionals Raise Usability Issues Differently Than Other Contributors?}

\subsubsection{Usability Dimensions}

\begin{figure}[t]
  \centering
    \includegraphics[width=\columnwidth]{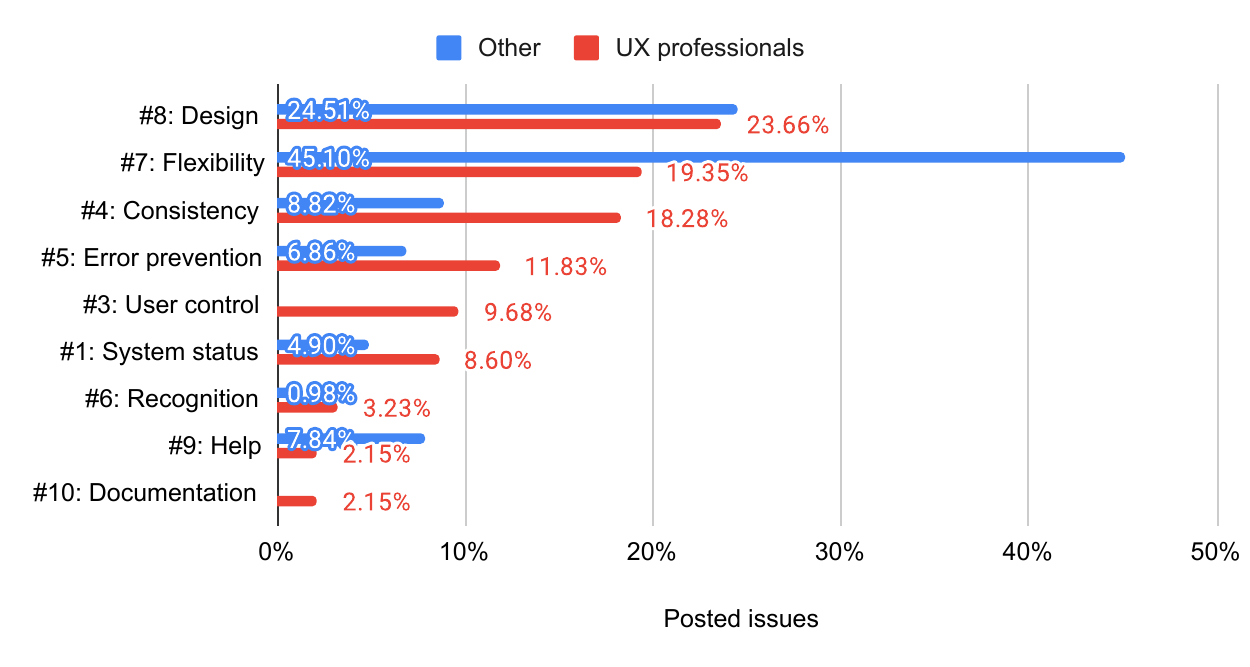}
    \caption{Percentage of issues posted by UX professionals and other contributors touching on different usability dimensions categorized by Nielsen heuristics.}
    \Description{A bar chart where the x-axis is the percentage and the y-axis includes nine usability heuristics. For UX professionals, 23.66 percent of the posted issues were related to heuristic \#8 Design, 19.35 percent related to heuristic \#7 Flexibility, 18.28 percent related to heuristic \#4 Consistency, 11.83 percent to \#5 Error Prevention, 9.68 percent to \#3 User control, 8.60 percent to \#1 System status, 3.23 percent to \#6, 2.15 percent to \#9, and 2.15 percent to \#10. For other contributors, 24.51 percent of the posted issues were related to heuristic \#8 Design, 40.10 percent related to heuristic \#7 Flexibility, 8.82 percent related to heuristic \#4 Consistency, 6.86 percent to \#5 Error Prevention, 4.90 percent to \#1 System status, 0.98 percent to \#6, and 7.84 percent to \#9.}
    \label{fig:frequency-Nielsen-heuristics}
\end{figure}

The usability issues posted by UX professionals reported a diverse range of usability concerns; see Figure~\ref{fig:frequency-Nielsen-heuristics}. We found that UX professionals considered most of the usability dimensions, covering a wider and more balanced range than other contributors. Similar to other contributors, UX professionals paid attention to \textit{\#7: Flexibility and efficiency of use} and \textit{\#8: Aesthetic and minimalist design}. However, their primary focus was also on \textit{\#4: Consistency and standards} and \textit{\#5: Error prevention}, while issues related to \textit{\#9: Help users recognize, diagnose, and recover from errors} were comparatively less frequent.

\begin{figure}[t]
  \centering
  \begin{minipage}
    [b]{\columnwidth}
    \includegraphics[width=\columnwidth]{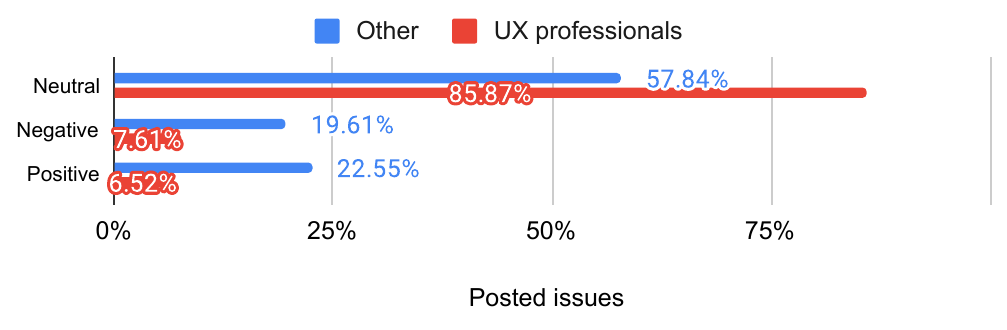}
    \subcaption{Sentiments}
    \Description{A bar chart where the x-axis is the percentage and the y-axis includes three sentiments: neutral, negative, and positive. For UX professionals, 85.87 percent of the posted issues were neutral, 7.61 percent were negative, and 6.52 percent were positive. For other contributors, 57.84 percent of the posted issues were neutral, 19.61 percent were negative, and 22.55 percent were positive.}
    \label{fig:frequency-sentiment-compare}
  \end{minipage}
  \hfill
  \begin{minipage}[b]{\columnwidth}
    \includegraphics[width=\columnwidth]{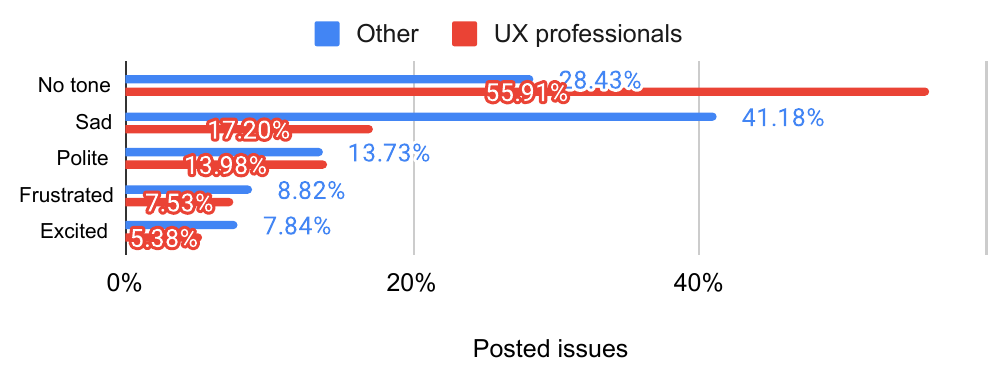}
    \subcaption{Tones} 
    \Description{A bar chart where the x-axis is the percentage and the y-axis includes five tones: sad, polite, frustrated, excited, and no tone. For UX professionals, 55.91 percent of the posted issues had no tone, 17.20 percent were sad, 13.98 percent were polite, 7.53 percent were frustrated, and 5.38 percent were excited. For other contributors, 28.43 percent of the posted issues had no tone, 41.18 percent were sad, 13.73 percent were polite, 8.82 percent were frustrated, and 7.84 percent were excited.}
    \label{fig:frequency-tone-compare}
  \end{minipage}
   \caption{Percentage of issues posted by UX professionals and other contributors that included different sentiments and tones.}
    \label{fig:frequency-emotion}
\end{figure}

\subsubsection{Sentiment and Tones}
The UX professionals more frequently applied \textit{neutral} sentiment and \textit{no tone} in comparison to others (see Figures~\ref{fig:frequency-sentiment-compare} and~\ref{fig:frequency-tone-compare}). Notably, other contributors frequently used the \textit{sad} tone when posting usability issues. This means that issues posted by UX professionals were more factual than emotional. This aligns with our impression when reading those issue posts.

\begin{figure}[t]
  \centering
    \includegraphics[width=\columnwidth]{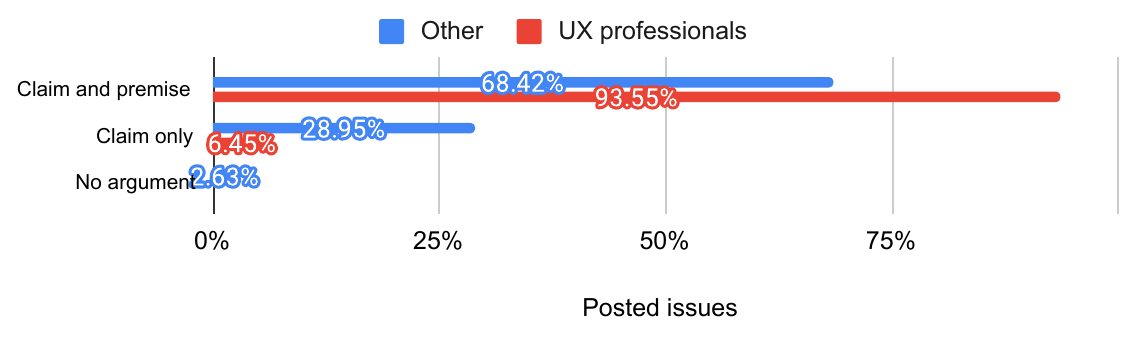}
    \caption{Percentage of issues posted by UX professionals and other contributors that included a claim and a premise.}
    \Description{A bar chart where the x-axis is the percentage and the y-axis includes three types: claim and premise, claim only, and no argument. For UX professionals, 93.55 percent of the posted issues had both claim and premise and 6.45 percent had only claim. For other contributors, 68.42 percent of the posted issues had both claim and premise, 28.95 percent had only claim, and 2.63 percent were not argumentative.}
    \label{fig:frequency-argument-structure}
\end{figure}

\subsubsection{Argument Structure}
Usability issues posted by UX professionals often had a solid premise to support their argument and they frequently adopted an argumentative structure of \textit{claim and premise}, as depicted in Figure~\ref{fig:frequency-argument-structure}. This is different than the other contributors, who more frequently reported issues without any premise.

\subsection{RQ2: How Do UX Professionals Follow Up on the Usability Issues They Posted?}

\subsubsection{Frequency of Follow-Ups}

Among the 93 usability issues posted by the four UX professionals, we found that they followed up on 31 (33.3\%) of them in the issue comments. In the remaining 62 issues that they did not follow up with, we found that most were resolved right after the issue post or at most after a few comments.

Focusing on UX professionals' general following-up behavior as commenters, some interesting observations appeared from our preliminary analysis. First, none of the UX professionals participated with other UX professionals; they only collaborated on the usability discussion threads that they instantiated. Additionally, all the \textit{claim-only} issues were not followed up by the UX professionals. Given the limited sample size of our dataset, these preliminary observations need to be further investigated in future work.

\subsubsection{UX Professionals' Purposes for Following up on Their Usability Issues}
\begin{figure*}[t]
    \centering
    \includegraphics[width=\textwidth]{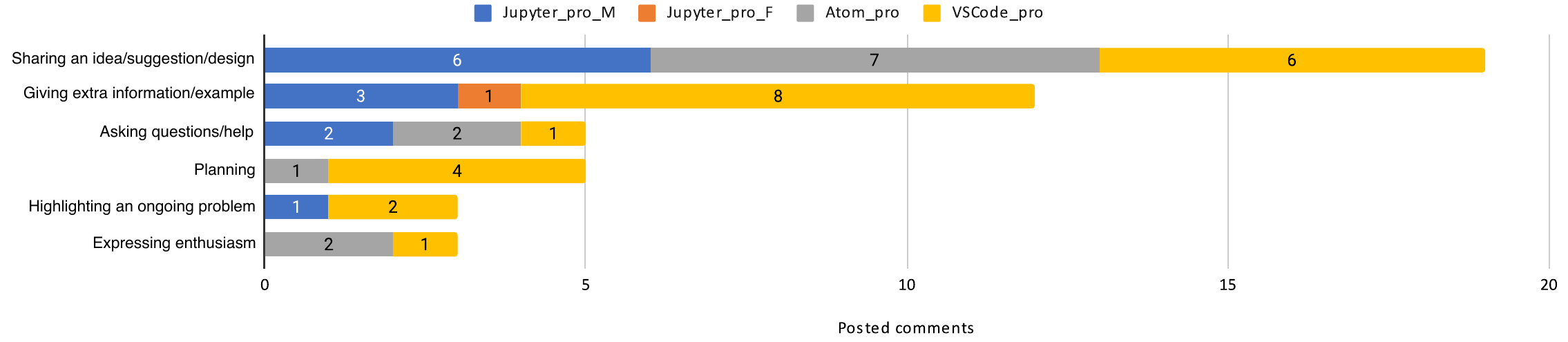}
    \caption{Frequency of UX professionals' purposes for following up on their usability issues}
    \Description{A bar chart where the x-axis is the frequency of posted comments and the y-axis includes the six purposes of following up. For "sharing an idea/suggestion/design", the frequency was 19; for "giving extra information/example", the frequency was 12; for "asking questions/help", the frequency was 5; for "planning", the frequency was 5; for "highlighting an ongoing problem", the frequency was 3; and for "expressing enthusiasm", the frequency was 3.}
    \label{fig:frequency-purpose}
\end{figure*}

Through the inductive coding process, we identified the following purposes of the four UX professionals when commenting in discussion threads about their usability issues; Figure~\ref{fig:frequency-purpose} shows the frequency of these purposes.

\addvspace{4pt}
\textbf{Sharing an idea/suggestion/design:} These commenters offer insights, suggestions, personal experiences, opinions, or design recommendations related to the topic under discussion. They contribute creative or evaluative input. One instance of this purpose is from Atom's UX professional in issue \#1006:

\begin{quote}
    \textit{Awesome, my real secret was I was waiting for @benogle's thoughts to go forward :) Next matter to discuss regarding these tabs is... Separate tabs from UI \& Syntax Styles: I've been noticing in my discovery that many themes have dark tabs/syntax and a light theme around the web. Others may prefer lighter UI\/Tabs with a dark UI. One thought would be to make them separate from the UI and the Syntax. For myself I'd match them up `usually' with the syntax and not the UI. Anyone with strong thoughts on this?}
\end{quote}

As another example, one of VSCode's UX professionals commented in issue \#18132: 

\begin{quote}
    \textit{@sandy081 nice work. As far as the title for the default settings, I vote that we leave it out unless we have indications from users that it's needed (perhaps we could try it first without it). However, here is an idea for a dismissible header. IMAGE. Here was another read-only concept I had (but I think the darker background is more clear). Just showing this as another example. I think we could do without it, since you're right it might raise questions on other editors that are read only.}
\end{quote}

\addvspace{4pt}
\textbf{Giving extra information/example:} The participants of comments contributed additional details, examples, or context to enrich the discussion. Their aim was to provide supplementary information or examples for a better understanding of a discussed topic. Examples of this type of purpose are VSCode UX professionals in issue \#4331: ``\textit{This will also be relevant to \#4100}'' and the male UX professional in Jupyter Lab of \#6615:

\begin{quote}
    \textit{@ellisonbg the folder icon is rendering in the correct alignment on my JupyterLab. But I haven't updated anything since yesterday; maybe it's a regression? I'm on Chrome {Version 74.0.3729.169 (Official Build) (64-bit)} and the latest OS.}
\end{quote}

\addvspace{4pt}
\textbf{Asking questions/help:} The comment posters sought clarification or assistance by posing queries or requesting guidance related to the posted issues. They aimed to gather information, guidance, or solutions. For example, Atom's UX professionals in issue \#964: ``\textit{What will it take to make this happen? Are we talking styles to move the drawer or more than that?}'' Also, Jupyter Lab male UX professionals in issue \#7967:

\begin{quote}
    \textit{I had a couple of questions about the existing UI. Is anybody attached to the @ symbols? They are a bit repetitive, I think we can go forward without them, but if somebody has a good reason to leave them in I am unaware of please let me know. Some of the extensions in the existing UI don't have orgs or usernames (see screenshot). Is this something the extension developer is doing on purpose?}
\end{quote}

\addvspace{4pt}
\textbf{Planning:} The UX professionals also discussed in the comments future strategies, proposed plans, or outlined potential courses of action to address the issue or improve the situation. They engage in forward-thinking discussions. One example of this type is from VSCode's UX professional in issue \#9861: ``\textit{Feel free to move to August if it's not worth the risk fixing it quickly.}'' and ``\textit{@isidorn FYI - this is the task I'll work on once your change is in Master. I'll put this on the June Milestone.}''

\addvspace{4pt}
\textbf{Highlighting an ongoing problem:} In these comments, the UX professionals drew attention to persistent issues, emphasizing the need for resolution or further investigation. They aimed to underscore existing problems for collective awareness and action. As an example for this type of purpose, a VSCode UX professional commented in issue \#3682: ``\textit{[IMAGE] This is me web inspecting it so you can see the problem (which is otherwise harder to see).}''

\addvspace{4pt}
\textbf{Expressing enthusiasm:} The commenters sometimes exhibited excitement, positive feedback, or encouragement regarding the topic or solution being discussed. Their goal in the comment was to express support or appreciation. One example of this is from Atom's UX professionals in issue \#840: ``\textit{Look forward to updating to check this out.}''

\section{Discussion}
We focused on analyzing the characteristics of the usability issues posted by UX professionals and understanding how they followed up on their usability issues on the ITSs. In this section, we first synthesize our findings related to the participation of UX professionals in the usability issue posting and discussions, and then we discuss the implications of these findings.

\subsection{Synthesizing the Primary Findings}

\subsubsection{Scarcity of UX Professionals' Engagement in OSS Usability Issue Discussions}
Our results yielded crucial evidence that UX professionals' involvement in the OSS issue discussions is very scarce; i.e., only four individuals were identified as UX professionals from the 224 usability issue posters. This limited presence raises concerns about the potential impact on UX quality and overall user satisfaction in OSS projects. We recognized that one possible reason behind their limited involvement in the issue tracking system is that they may conduct and discuss their work related to usability, such as user studies and UX design, outside of these systems. However, we argue that because the central development decisions, project management, and collaboration for OSS take place within ITSs, the limited participation of UX professionals on these platforms creates a potential disconnection between the design considerations that they offer and the actual development workflow. This disconnection is also problematic as it may lead to a lack of alignment between design goals, overall project objectives, and lack of user satisfaction. Thus, while we recognize that UX and usability-related tasks go beyond issue discussions, understanding the implications of limited contribution by UX professionals in these systems is essential for addressing challenges and fostering more effective collaboration in OSS development.

\subsubsection{Specific Issue Reporting and Follow-Up Style of UX Professionals}
A fascinating aspect of UX professionals' contributions was their distinct issue-reporting style and issue follow-up behavior. Usability issues posted by UX professionals covered a wider range of usability aspects than those posted by other contributors. UX professionals also almost always backed up their usability-related arguments with premises and tended to approach usability issues using their expertise, without leveraging emotional tone or sentiment. Moreover, when following up on the issues they posted, the UX professionals often made efforts to ensure that their points were clearly understood and their concerns were properly addressed. These findings all indicated that the expertise of the UX professionals indeed affected the way they posted usability issues. The issues and comments posted by the UX professionals read confident, well-explained, and sufficiently supported. These attributes may help raise the awareness of usability in the OSS communities, allowing them to adopt a user-centric mindset and better prioritize these issues. Together, our findings highlighted the multifaceted nature of UX professionals' contributions to usability issue discussions within OSS projects, although their current involvement is rare. These insights may provide directions for further investigation and potential strategies for enhancing the collaboration and engagement of UX professionals in OSS communities in the future.

\subsection{Implication to Practice and Research}
Our results about the actual contributions provided by UX professionals carry important implications for both OSS practices and future research. According to our study's insights into UX professionals' distinctive reporting styles, developing collaboration tools tailored to their communication patterns could support more effective collaboration within OSS communities. For instance, platforms that encourage factual and experience-based reporting, frequently adopted by the UX professionals in our dataset, may enhance their involvement and further benefit other stakeholders (e.g., end users) in creating usability issues. These features can also enhance issue summarization techniques (e.g.,~\cite{Gilmer2023}) to support collaborative issue comprehension and synthesis. Moreover, designing tools that match UX designers with OSS projects based on the designers' backgrounds and skills, as well as the OSS projects' characteristics, could encourage both project maintainers to actively seek support from UX professionals and UX professionals to contribute to OSS projects. Besides, tools could be designed to identify specific usability issue types (e.g., based on usability heuristics) and, depending on the issue type, help UX professionals apply their expertise more effectively in discussing and resolving the issues.

Further, understanding how UX professionals posted and discussed usability allows us to envision collaboration strategies that can maximize their involvement within OSS projects. Integrating their expertise in the entire development life-cycle, including expert review, user testing, and design would offer opportunities to shape the project direction from the beginning. Dedicated spaces for UX-focused discussions on OSS products would facilitate knowledge and experience sharing as well as collective problem-solving among UX professionals. Besides, incorporating UX-related metrics in OSS platforms such as GitHub would highlight the influence of UX contributors, reinforcing the value of their role within the OSS communities.

\subsection{Limitations and Future Work}
This research suffers from multiple limitations that can be addressed in future work. First, we recognize that this research analyzed only a few usability issues reported by UX professionals and other contributors based on an existing dataset from previous research~\cite{sanei2023characterizing}. To enhance generalizability, future studies should aim to expand the sample of OSS usability issues. Similarly, this study only investigated a few popular OSS projects. Our rationale was that popular projects may attract more UX professional contributions. However, smaller projects may have unique characteristics that we did not capture. Further, we did not investigate collaboration platforms other than the ITS (GitHub Issues in this case). We chose this platform since it is the place where most of the public decision-making of OSS is happening. Yet, we acknowledge that some UX discussions may occur on other platforms, including private channels of the projects that we cannot observe. Finally, this research is based on an analysis of artifacts created by OSS community members. Although this method provided rich information about the main focuses of the OSS communities regarding UX professionals in reporting usability issues, it cannot reveal personal cognitive aspects such as goals, motivations, and challenges of usability issue posters. Therefore, by conducting direct user studies with UX professionals who contribute to the OSS communities, future work could have a richer understanding of their roles in OSS.

\section{Conclusion}
This study focused on a preliminary investigation of the characteristics of usability-related issues and comments posted by UX professionals in OSS issue tracking systems. The results indicated a scarcity in the participation of this group of contributors within such a platform. However, issues posted by UX professionals displayed more complete arguments and unemotional features, highlighting the positive influence of the expertise possessed by these contributors. Our results provided useful insights for facilitating the involvement and collaboration of UX professionals to enhance the usability and design quality of OSS.

\begin{acks}
This work is partially supported by the Alfred P. Sloan Foundation (G-2021-16745) and the Natural Sciences and Engineering Research Council of Canada (RGPIN-2018-04470).
\end{acks}

\balance
\bibliographystyle{ACM-Reference-Format}
\bibliography{references}

\end{document}